\documentstyle[psfig,11pt,aaspp4,]{article}

\received{}
\accepted{}
\journalid{}{}
\articleid{}{}

\slugcomment{Accepted by {\it The Astrophysical Journal}}

\lefthead{Halpern {\it et al.}}
\righthead{Possible Counterpart of 3EG J2227+6122}

\begin{document}

%
%
\def\lsim{\mathrel{\lower .85ex\hbox{\rlap{$\sim$}\raise
.95ex\hbox{$<$} }}}
\def\gsim{\mathrel{\lower .80ex\hbox{\rlap{$\sim$}\raise
.90ex\hbox{$>$} }}}

\def\source{3EG~J2227+6122}
\def\ro{{\it ROSAT\/}}
\def\asca{{\it ASCA\/}}
\def\mystery{RX~J2229.0+6114}
\def\xray{RX/AX~J2229.0+6114}
\def\vla{VLA~J2229.0+6114}

\title{A Possible X-ray and Radio Counterpart of the High-Energy
Gamma-ray Source 3EG J2227+6122}

\author{J. P. Halpern\altaffilmark{1},  E. V. Gotthelf, K. M. Leighly,
and D. J. Helfand}
\affil{Columbia Astrophysics Laboratory, Columbia University,
550 West 120th Street, New York, NY 10027}
\authoremail{jules@astro.columbia.edu}
\altaffiltext{1}{Visiting Astronomer, Kitt Peak National Observatory,
National Optical Astronomy Observatories, which is operated by 
the Association of Universities for Research in Astronomy, Inc. (AURA) 
under cooperative agreement with the National Science Foundation.}

\begin{abstract}
\rightskip 0pt \pretolerance=100 \noindent
The identity of the persistent EGRET sources
in the Galactic plane is largely a mystery.  For one of these,
\source, our complete census of X-ray and radio sources
in its error circle reveals a remarkable superposition of
an incomplete radio shell with
a flat radio spectrum, and a compact,
power-law X-ray source with photon index $\Gamma = 1.5$
and with no obvious optical counterpart.  The radio shell is
polarized at a level of $\simeq 25\%$.
The anomalous properties of the radio source prevent us
from deriving a completely satisfactory theory as to its nature.
Nevertheless, using data from \ro, \asca, the VLA,
and optical imaging and spectroscopy,
we argue that the X-ray source may be a young pulsar with an associated
wind-blown bubble or bow shock nebula,
and an example of the class of radio-quiet 
pulsars which are hypothesized to comprise the majority of EGRET
sources in the Galaxy.  The distance to this source can
be estimated from its X-ray absorption as 3~kpc.  At this
distance, the X-ray and $\gamma$-ray luminosities would be
$\approx 1.7 \times 10^{33}$ and $\approx 3.7 \times 10^{35}$
erg~s$^{-1}$, respectively, which would require an
energetic pulsar to power them. 

If, on the contrary, this X-ray source is {\it not} the
counterpart
of \source, then by process of elimination
the X-ray luminosity of the latter must
be less than $10^{-4}$ of its $\gamma$-ray luminosity,
a condition not satisfied by any established class of $\gamma$-ray
source counterpart.  This would require the existence of
at least a quantitatively new type of EGRET source,
as has been suggested in studies of other EGRET fields.
\end{abstract}

\section{Introduction}

The nature of the persistent high-energy ($> 100$ MeV)
$\gamma$-ray sources in the Galaxy remains an enigma almost
three decades after their discovery.
The third installment of the EGRET catalog
(Hartman et al. 1999) lists a total of 270 sources, of which 93 are
likely or possibly identified with blazars,
five with rotation-powered pulsars, one with Cen~A,
and one with the LMC. This leaves 170 unidentified sources,
of which 57, or one third, lie within
$|b| \leq 10^{\circ}$ along the Galactic plane.  This excess
of low-latitude sources must comprise a
Galactic population that is either similar to the already
identified $\gamma$-ray pulsars, or representative of an entirely new class
of $\gamma$-ray emitter associated with the disk population.
These Galactic sources have proven extremely difficult
to identify.    

Rotation-powered pulsars are likely to dominate
the Galactic $\gamma$-ray source population, but their
detectability at both radio and $\gamma$-ray wavelengths
depends on their beam patterns.
The shapes of radio pulsar beams as determined by the
rotating vector model (Radhakrishnan \& Cooke 1969)
demands that $\sim 50-70\%$ of young radio pulsars
are {\it not} visible from Earth (Frail \& Moffett 1993;
Tauris \& Manchester 1998).  The clear differences between
the broad observed $\gamma$-ray beam patterns and the narrow radio pulses
implies that $\gamma$-ray emission is probably visible from
a more complete range of directions than are the radio beams.
Indeed, the identification of the high-energy $\gamma$-ray source
Geminga as the first radio quiet,
but otherwise ordinary pulsar (Halpern \& Holt 1992; Bertsch et al. 1992),
provides a likely prototype for the remaining unidentified Galactic sources.
That Geminga was the brightest unidentified
EGRET source, at a distance of
only $\sim 250$~pc, argues that there should be others.
The possible existence of intrinsically radio-quiet
$\gamma$-ray pulsars also cannot be ruled out.

Several authors have considered the pulsar hypothesis from statistical
or theoretical points of view (Halpern \& Ruderman 1993;
Helfand 1994; Kaaret \& Cottam 1996).
The most detailed theoretical treatment of the pulsar model for
the Galactic $\gamma$-ray sources is that of Romani \& Yadigaroglu (1995)
and Yadigaroglu \& Romani (1995;1997).
They developed a numerical calculation of $\gamma$-ray production
and beaming in the outer-gap model that
successfully reproduces the
basic observed features of the pulse profiles and the $\gamma$-ray
efficiency as a function of age.  By combining this model with a
Monte Carlo simulation of the Galactic pulsar population,
they estimated that a total of 22 pulsars should be detected
by EGRET at the threshold of the first EGRET catalog.  This number
is very close to the actual number of unidentified sources
at $|b| < 10^{\circ}$ in that catalog.

Candidate neutron-star identifications for additional EGRET sources
have been found in the form of point-like X-ray sources with
no obvious optical counterparts.  One such candidate is
present in the supernova remnant CTA1 which is close to
3EG~J0010+7309 (Lamb \& Macomb 1997; Brazier et al. 1998).
Another one is in the $\gamma$-Cygni supernova remnant which is
coincident with 3EG~J2020+4017 (Brazier et al. 1996).  Both of these
so far lack detected of pulsations.  Most notable among the
probable identifications is the radio star
and Be/X-ray binary LSI~$+61^{\circ}303$ (Strickman et al. 1998)
which has long been
associated with the $\gamma$-ray source 2CG~135+01.
Another possible candidate
is a 34~ms X-ray pulsar with a Be star companion 
in the error circle of 3EG~J0634+0521 (Kaaret et al. 1999;
Cusumano et al. 2000).
Additional EGRET sources have recently been tentatively identified
with known radio pulsars
based on the detection of a corresponding pulsed $\gamma$-ray signal, 
namely 3EG~J1048--5840 with PSR B1046--58 (Kaspi et al. 2000),
and the millisecond pulsar PSR~J0218+4232 with 3EG~J0222+4253
(Kuiper et al. 2000).   X-ray synchrotron nebulae that are inferred
to be powered by pulsars have been detected in the error circles of
the EGRET sources 2EG~J1811--2339 (Oka et al. 1999) and
3EG~J1420--6038 (Roberts \& Romani 1998; Roberts et al. 1999).
PSR~B1951+32 in the supernova remnant CTB~80 has been detected
by EGRET (Ramanamurthy et al. 1995) even though it does not exceed
the threshold for inclusion in the EGRET catalogs.  It has also been
suggested that accreting neutron-star binaries might be EGRET sources,
although so far only one example
has been found in the possible detection of intermittent $\gamma$-rays
from Cen X-3 (Vestrand, Sreekumar, \& Mori 1997).

Informed by the pulsar scenario, we are studying several EGRET sources
that are at ``intermediate'' Galactic latitudes,
$3^{\circ} < |b| < 8^{\circ}$, and that are not apparently variable.
This strategy increases the likelihood that ({\it a}) the 
EGRET source is real, 
({\it b}) its position is not affected by errors in the diffuse
emission model or neighboring weak sources, ({\it c})
it is relatively nearby, ({\it d}) the intervening column density is not
too large for soft X-ray observations, and ({\it e}) the corresponding
optical fields are not too crowded.
The absence of variability is important, since the known $\gamma$-ray
pulsars show little if any long-term variability, while the blazars often
flare dramatically.

One of our targets is \source, a source at Galactic coordinates
$(\ell,b) = (106.\!^{\circ}5,3.\!^{\circ}2)$ with a 95\%
error circle of radius $0.\!^{\circ}46$ (Hartman et al. 1999).
Its average flux is $4.1 \times 10^{-7}$ photon~cm$^{-2}$~s$^{-1}$
($> 100$~MeV)
and its photon spectral index is $2.24 \pm 0.14$.  It is not obviously
variable (McLaughlin et al. 1996).  The total Galactic 21 cm column
density in this direction is $8.2 \times 10^{21}$~cm$^{-2}$
(Stark et al. 1992).  Prior to this work, there were no known
pulsars or blazar-like radio sources in this field, and no prominent
X-ray sources.  In this paper we present the results of X-ray, radio,
and optical observations of the region of \source\
which together point to a possible identification.

\section{Observations}

\subsection{\ro\ HRI Survey}

The \ro\ HRI was used to cover the 95\% error circle
of \source\ in four overlapping pointings, as shown
in Figure~1.  These observations were made during
1996 August 7--13.  Exposure times ranged between
14,000 and 19,000~s.  A total of six compact X-ray sources
were detected in this field to a limiting count rate of approximately
$1 \times 10^{-3}$ counts~s$^{-1}$.   This limit corresponds to
an intrinsic flux of $4 \times 10^{-14}$ erg~cm$^{-2}$~s$^{-1}$
for a thermal plasma of $T = 3 \times 10^6$~K and
$N_{\rm H} = 2 \times 10^{20}$~cm$^{-2}$,
spectral parameters that are expected of stellar coronal
X-ray sources which should be the dominant population in this field.
The X-ray source positions, count rates, and optical counterpart data
are listed in Table~1.  Optical magnitudes 
and positions are taken from the USNO--A2.0
catalog (Monet et al. 1996).

\subsection{Optical Identifications of \ro\ Sources}

Optical spectroscopic identifications were made for five
of the six \ro\ HRI sources using the KPNO 2.1m telescope and
Goldcam spectrograph.  Four of these sources are bright K and M
type stars.  The fifth is a bright emission-line star which is
classified as a Herbig Ae/Be type by Hang, Liu, \& Xia (1999);
it is also an {\it IRAS} source.  Our optical spectrum of this star
is essentially identical to that of Hang et al., 
and it confirms their classification in detail.  This star also illuminates
a prominent nebulosity which is visible on the Palomar
Observatory Sky Survey (POSS) plates.  Our spectroscopy and subsequent
H$\alpha$ imaging show that this nebula is bright in H$\alpha$.
Herbig Ae/Be stars are commonly detected as X-ray sources
(e.g., Zinnecker \& Preibisch 1994), but we have no reason to suspect
that this one is the origin of \source.  We also note that the 
X-ray positions of all five identified sources are coincident
with their optical positions to the expected statistical accuracy of 
$1^{\prime\prime}-3^{\prime\prime}$ (see Figure~2).  Furthermore,
any systematic offset between the average X-ray 
and optical positions is $\leq 0.\!^{\prime\prime}6$,
which is not significantly different from zero. Therefore,
we do not find it necessary to make any {\it a posteriori}
adjustments to the X-ray astrometry.

The sixth HRI source, \mystery, remains unidentified optically.
The remainder of this paper is largely concerned with the properties
of this source and the evaluation of its credentials as a possible
identification of \source.
Within a conservative $5^{\prime\prime}$ error radius, 
which is justified empirically by the data in Figure~2, there
is nothing at this location on the POSS plates.
Figure~3 shows images of this
field to a limit of $R=24.5$ and $B=24.0$ that we obtained
using the MDM 2.4m telescope.  The nearest
bright object is a star of $R \approx 16.8$ 
which lies $5.\!^{\prime\prime}6$ southeast of the X-ray
position; it is labeled star~A in Figure~3.
An optical spectrum which we obtained on the KPNO 2.1m telescope
shows that it is in fact a highly reddened A~star, an unlikely
counterpart considering the hard X-ray properties of this
source discussed below.  The brightest object
in the X-ray error circle, just north of star~A, has $R = 21.3$.
A Keck spectrum of this object kindly obtained by
R.~H. Becker shows no emission lines or other interesting features.  
Although we cannot firmly classify the spectrum, it is possibly
an ordinary star or galaxy.  On the other hand, if it is a truly
featureless spectrum, then it could be a Crab-like pulsar or a
BL~Lac object.  Nonetheless, we regard this X-ray source as
unidentified to a limit of $R\geq 21.3$ and $B>24$, and possibly to
$R \geq 23$ if the $R = 21.3$ object is not its counterpart.

\subsection{Radio Observations}

We created a 20 cm image of the entire EGRET error circle for \source\ by
constructing a mosaic from 13 snapshots obtained on 31 March 1996 using
the Very Large Array (VLA\footnote{The VLA, part of the National Radio Astronomy
Observatory, is operated by Associated Universities, Inc., under cooperative
agreement with the National Science Foundation.}) in its C configuration.
The synthesized beam yields a resolution of $\approx 15^{\prime\prime}$
FWHM; the median rms in the image is 0.12~mJy. Subsequently, the NRAO/VLA Sky
Survey  (NVSS -- Condon et al. 1998) covered this field to a somewhat lower
sensitivity (rms $\simeq 0.5$~mJy) at an angular resolution of 
$45^{\prime\prime}$.  In addition, we have examined
images from the 92~cm Westerbork Northern Sky Survey 
(WENNS -- Rengelink et al. 
1997), the Greenbank 20~cm single-dish image (Condon \& Broderick 1991),
and the Greenbank 6 cm catalog (Becker, White, \& Edwards 1991),
as well as observing a portion of the field again at 6~cm
on 12 January 1997 with the VLA in its D configuration.

Only one radio source in the EGRET error circle is coincident with an X-ray
source: the single unidentified source \mystery.  An image of this
radio source constructed from the $I$, $Q$, and $U$ 
maps of the NVSS is displayed in Figure
4$a$ where it is seen that \mystery\ lies at the center of an incomplete circular
radio shell with a diameter of $3.\!^{\prime}5$. The NVSS catalog lists two 
components for this source with a combined total flux density of
$\approx 73 \pm 5$~mJy. More remarkable is the high degree of linear 
polarization present throughout the shell. The NVSS catalog lists a polarized flux
density of $\sim 21$~mJy, albeit with large reported errors.
However, we have constructed a polarization map from the archived $Q$ and $U$ data,
and the signal is unmistakable.
We find a peak polarized flux density of 3.9~mJy per beam (the map rms
is 0.23~mJy) and an integrated polarized flux density of 17 mJy, yielding 
a polarized fraction of $\simeq 25\%$. In contrast, the 20~mJy source 
$4^{\prime}$ south of the shell has a maximum polarized signal of 0.8~mJy,
yielding an upper limit to its
polarization of $\simeq 4\%$. Other sources in the
field are also unpolarized at the 5\% level. The polarization vectors are 
approximately radial throughout the shell, implying a tangential magnetic field.
However, it is important to note that an unknown amount of Faraday rotation 
along the line of sight to the source could alter this interpretation.

Our additional observations of this source at 20 and 6 cm confirm its high 
degree of polarization.  While our higher resolution data clearly
over-resolve the source, leading to substantial missing flux density,
both images also yield polarized fractions of $\simeq 30\%$;
the 6~cm image is displayed in Figure 4$b$.
The orientation of the polarization vectors at 6~cm
are similar to those in the NVSS image, although they are not reliable
since no polarization
calibration was carried out during this observation.  Note that the source
$4^{\prime}$ south of the shell is resolved in this map into an
elongated structure $\approx 20^{\prime\prime}$ in extent, reminiscent of an
extragalactic double radio source; its steep spectral index of
$\alpha = 0.9$ (where $S_{\nu} \propto \nu^{-\alpha}$)
derived from these data in conjunction with the WENNS catalog entry is
consistent with this interpretation.

No source is listed in the WENNS catalog at the position of radio shell.
However, a map extracted from the archive shows a clear excess coincident
with the shell; an image smoothed with a $60^{\prime\prime}$ Gaussian yields
a crude estimated flux density of $\sim 35$~mJy,
which is probably uncertain by a factor of 2.
It is clear from this image, however, that while the southern
source has peak and integrated flux densities of 51 mJy and 60 mJy, 
respectively, the shell source is not significantly brighter at 92~cm than
it is at 20~cm.

The shell source is not detected in the Greenbank 20~cm images because of 
confusion with bright diffuse emission nearby, but is clearly seen in the 
Greenbank 6~cm maps; its flux density is listed in Becker et al. (1991)
as 80~mJy. Since
the $3.\!^{\prime}5$ beam of the Greenbank telescope at this wavelength is
well-matched to the source size, this is probably the most reliable measure
of the source flux density, as all of
the interferometric measurements resolve out
some fraction of the flux. Our scaled-array observations with $\approx 15^{\prime\prime}$
beams yield the same flux densities at the two frequencies to within the 
relatively large errors. Thus, we conclude that all available measurements
are consistent with a flat spectral index for the radio shell
from 92 cm through 6 cm, with an
integrated flux density of $\approx 80$~mJy and a
$\simeq 25\%$ polarized fraction.

There are no other notable radio sources in the error circle of \source.
The brightest source, with a flux density at 6~cm of 494 mJy (Becker et al.
1991), is the H~II region Sharpless~141. Most significant for the classification
of the $\gamma$-ray source, however, is the fact that there is no compact,
flat-spectrum radio source in this field comparable to the well-identified
EGRET blazars, all of which have 6~cm flux densities in excess of 400~mJy
(Mattox et al. 1997). The upper limit on such a source in the field of
\source\ is $\simeq 20$ mJy, the flux density limit of the Becker et al. 
catalog.

\subsection{H$\alpha$ Imaging}

To search for further evidence concerning the nature of the radio nebula
and the compact X-ray source,
we obtained H$\alpha$ images of an $7.\!^{\prime}3 \times 7.\!^{\prime}3$
region around \vla\ using the MDM 2.4m telescope and a 39~\AA\
wide filter centered at 6563 \AA.  Figure~5 shows the combined image.
Diffuse H$\alpha$ structures are present throughout,
with a peak surface brightness of $1.7 \times 10^{-16}$
erg~cm$^{-2}$~s$^{-1}~{\rm arcsec}^{-2}$ above the background level
in the northwest and southeast corners of the image.
This value is comparable to the average of the diffuse ionized
gas at $b = 0^\circ$ near this location
($6.3 \times 10^{-17}$ erg~cm$^{-2}$~s$^{-1}~{\rm arcsec}^{-2}$;
Reynolds 1985).  However, there is no
structure that appears correlated with either the radio shell or the
location of the X-ray source.  The $1\sigma$ noise level
in this image is
$1.2 \times 10^{-17}$ erg~cm$^{-2}$~s$^{-1}~{\rm arcsec}^{-2}$;
the implications of the lack of H$\alpha$ emission at this level
from \vla\ or \mystery\ will be discussed in \S 3.

\subsection {\asca\ Observation of the Unidentified Source \mystery }

To investigate further the nature of \mystery, we
obtained an \asca\ observation beginning on 1999 August~4.
A total of 114,500~s of exposure was obtained with each of the two
GIS detectors, and 97,600~s with each of the two SIS detectors operated
in 1-CCD mode.
A prominent hard X-ray source was detected at the position
(J2000) $22^h29^m05.\!^s9,\ +61^{\circ}14^{\prime}16^{\prime\prime}$
(corrected for the \asca\ temperature variation by the method
of Gotthelf et al. 2000).  This position is consistent with that
of the \ro\ HRI source to within $8^{\prime\prime}$, and is well
within the $12^{\prime\prime}$ radius \asca\ error circle (at 90\%
confidence).
A contour map of the combined \asca\ GIS
images is superposed on the \ro\ HRI images in Figure~1,
and a more detailed view of the GIS image is shown in Figure~6.
Analysis of the SIS and GIS photons from \xray,
extracted using standard methods, shows that the spectrum (Figure~7)
is best fitted by a power law of photon index $\Gamma = 1.51 \pm 0.14$
at 90\% confidence, and $N_{\rm H} = (6.3 \pm 1.3) \times 10^{21}$~cm$^{-2}$;
the confidence contours of these spectral parameters are shown in Figure~8.
The intrinsic 2--10~keV flux is $1.56 \times 10^{-12}$~erg~cm$^{-2}$~s$^{-1}$.

Some care was required in the extraction of source and background regions
for this analysis since there is also diffuse X-ray flux in this region. 
Weak, diffuse emission surrounding the compact source as well 
as to the northwest of it appears to be much softer than the compact source,
and its uncertain distribution is a significant source of systematic error
in the spectral parameters of \xray. 
We chose a radius of $4^{\prime}$ for the source region in all detectors,
and background regions which are as close as possible, but which exclude a 
region of radius $5^{\prime}\!.25$ around the source.
As we change the source and background extraction
regions, the spectral index of \xray\ can become as steep as 1.8, which
we consider the maximum systematic deviation from the best fitted value
of $1.51 \pm 0.14$.
There is also a weak, soft \asca\
source coincident with the Herbig Ae/Be star RX~J2226+6113.
No other significant sources are seen in the ASCA images to a flux limit of
$\sim 6 \times 10^{-14}$~erg~cm$^{-2}$~s$^{-1}$.

The fitted $N_{\rm H} = (6.3 \pm 1.3) \times 10^{21}$~cm$^{-2}$
of \xray\ is almost equal to the total Galactic 21~cm
column in this direction, $N_{\rm H} = 8.4 \times 10^{21}$~cm$^{-2}$
(Stark et al. 1992), indicating that the X-ray source is
at least 2~kpc distant, and possibly much farther.  In the following
sections we adopt a fiducial distance of 3~kpc as typical for
the X-ray measured $N_{\rm H}$ if \xray\ is Galactic.  Absorption
accounts for the relative weakness of the source in the HRI.
The fitted \asca\ spectral parameters would extrapolate
to an HRI count rate of $(5.5 \pm 1.5) \times 10^{-3}$ counts~s$^{-1}$,
approximately twice that observed.  This difference can be taken
either as marginal evidence of long-term variability of this X-ray source,
or as an indication that some extended emission is escaping detection
by the HRI.

We also searched the \asca\ spectrum of \xray\ for an emission
line of Fe~K$\alpha$. No such line is detected, and
95\% confidence upper limits on its EW range from 300~eV for
a narrow line at $z = 0$, to 50~eV for a narrow line at $z = 0.12$.
For a broad line (Gaussian $\sigma = 0.5$~keV), the corresponding limits are
440~eV at $z = 0$ and 150~eV at $z = 0.12$.  Beyond $z = 0.12$,
the X-ray luminosity of the source would exceed $10^{44}$~erg~s$^{-1}$,
and such luminous AGNs do not usually exhibit an Fe~K$\alpha$ line.

The elapsed time of the \asca\ observation was 67~h.  During this
time there was no evidence for short-term variability on time scales
of hours.  The \asca\ GIS bit assignments were configured for high time
resolution in order to search for pulsations.  Approximately
54,600~s of data were obtained with 3.9~ms resolution, and
59,100~s with 0.5~ms resolution.  These were searched,
without success, for periodic
pulsations for all periods as short as 10~ms using the Rayleigh test.
In addition to a coherent search of the entire photon list (extracted
from a region of radius $4^{\prime}$ around \xray\ ),
the period search was performed
on individual segments of 30,000~s in length in case there is a
short-period pulsar with a large $\dot P$ or acceleration.
For example, the Crab pulsar's $\dot P$
term contributes --0.17 cycles over an elapsed time of 30,000~s.
Further searches of the data for a high $B$ pulsar are in progress, and the
results will be reported in a subsequent paper.

\section {Possible Interpretations of \xray\ and \vla\ }

\subsection {A Chance Coincidence?}

We considered the hypothesis that \xray\ is a chance
superposition of a Galactic or extragalactic X-ray source
unrelated to the radio shell \vla.
The X-ray properties of \xray\ are by themselves
inconclusive as to its Galactic or extragalactic origin.
Power-law fits to the \asca\ spectrum require a column density
slightly less than the maximum Galactic 21~cm value, which allows either
a background AGN or a distant Galactic source.
The power-law photon index $\Gamma = 1.5$ is harder than that of
most radio-quiet QSOs, but is typical for a $\gamma$-ray pulsar
(Wang et al. 1998).
Although there is no prominent Fe~K$\alpha$ emission
line as would be expected from a low-redshift Seyfert galaxy,
we cannot rule out that this is a luminous QSO for which
Fe~K$\alpha$ is generally not seen. 
The absence
of variability on short and long time scales is also ambiguous,
since AGNs are often but not always variable during an observation
of this length.  The spatial resolution and sensitivity 
of the \asca\ SIS are not adequate to determine if 
\xray\ is spatially extended, which if true, would be strong 
evidence of a Galactic pulsar-powered nebula.  Similarly, it
is not possible to conclude whether or not the diffuse soft X-ray
emission surrounding \xray\ is associated with it.

The absence of an emission-line
optical counterpart to a limit of $R > 23$ somewhat favors 
a neutron star,
although the Galactic absorption $A_R = 4.6$~mag 
(Schlegel et al. 1998) allows for the
possibility that a faint object in the error circle in Figure~3
is a QSO of dereddened $R \approx 18.4$.
Such a QSO would
have $\alpha_{\rm ox} \approx 1.0$, which would be near the extreme
of X-ray loudness among QSOs (Wilkes et al. 1994).  Thus, we regard
the absence of a suitable optical counterpart for
\xray\ as somewhat discouraging an AGN classification unless
it is a new type of extreme $\gamma$-ray quasar (see \S 4).

\subsection {An H~II Region?}

We also considered the hypothesis that, if the X-ray source
\mystery\ is unrelated to the radio shell, then the latter
might be an H~II region.  The principle reason for entertaining
this possibility is the flat radio spectrum of \vla,
which is reminiscent of thermal bremsstrahlung
for which $S_{\nu} \propto \nu^{-0.1}$ in the radio.
While the high degree of polarization strongly favors
nonthermal emission, we also have independent evidence 
against the H~II region hypothesis from our optical
imaging.  The argument proceeds as follows. We can use the observed
radio flux to estimate the required electron density and
ionizing flux.  The thermal bremsstrahlung emissivity is
$$\epsilon_{\nu}\ =\ 6.8 \times 10^{-38}\ {Z^2 n_e n_i \over T^{1/2}}\
e^{-h\nu/kT}\ g_{ff}(\nu,T)\ \ {\rm erg\ cm^{-3}\ s^{-1}\ Hz^{-1}}.
\eqno(1)$$
Assuming $T = 8,000$~K and an observing frequency $\nu = 4.86 \times
10^9$~Hz, the Gaunt factor $g_{ff} \approx 4.91$.  Approximating the
nebula as a sphere of ionized hydrogen of radius $r_s = \theta d/2$ where
$\theta \approx 3^{\prime}$ is its angular diameter,
the flux density received at Earth is
$$S_{\nu}\ =\ 0.095\ \left  ({d \over 3\,{\rm kpc}} \right )\ n_e^2\
\ {\rm mJy}.\eqno(2)$$
In this case, since we observe $S_{\nu} \approx 80$~mJy,
the mean electron density in the nebula must be
$n_e \approx 30$~cm$^{-3}$.  

The rate of ionizations $Q$ is related to the Str\"omgren radius
$r_s$ and the case B recombination coefficient $\alpha_B(T)
\approx 3.1 \times 10^{-13}$ cm$^3$~s$^{-1}$ via
$$Q\ =\ {4\pi \over 3}\ r_s^3\ n_e\,n_+\ \alpha_B(T)\ 
=\ 8.4 \times 10^{43}\ \left ({d \over 3\,{\rm kpc}} \right )^3\ n_e^2\ \
{\rm s^{-1}}.\eqno(3)$$
Combining Equations (2) and (3), we can eliminate $n_e$
and solve for $Q$ as function of
the observed radio flux and the unknown distance $d$,
$$Q\ =\ 7.1 \times 10^{46}\ \left ({d \over 3\,{\rm kpc}} \right )^2\
\left ({S_{\nu} \over 80\,{\rm mJy}} \right )\ \ {\rm s^{-1}}.\eqno(4)$$
For a range of plausible distances between 2 and 8~kpc, $Q$ is in the
range $(0.32-5.0) \times 10^{47}$~s$^{-1}$, which corresponds to
stars of spectral type B0.  The absolute visual
magnitudes of such stars are in the range --4.1 to --4.4.  Taking
into account an estimated extinction $A_V = 1.4$ mag~kpc$^{-1}$
(up to a maximum of $A_V = 5.7$), the apparent magnitude of the
exciting star should fall in the range $m_V = 10.3-15.8$.

The main difficulties with the H~II region hypothesis are the absence of
H$\alpha$ recombination radiation and the lack of a suitable exciting
star.  We describe each of these failed predictions in turn.
Corresponding to the radio bremsstrahlung emission there should be
H$\alpha$ with emission coefficient
$\epsilon({\rm H}\alpha)\ =\ 3.2 \times 10^{-25}$ erg~cm$^3$~s$^{-1}$,
giving a total observed flux of 
$$F({\rm H}\alpha)\ =\  8.2 \times 10^{-14}\ 
\left ({d \over 3\,{\rm kpc}} \right )\ n_e^2\ \
{\rm erg\ cm^{-2}\ s^{-1}}.\eqno(5)$$
It is more useful to relate
the expected surface brightness in H$\alpha$ directly
to the surface brightness in the radio image by combining Equations (2)
and (5).  Independent of any other variables, the proportionality is
$$\Phi({\rm H}\alpha)\ =\ 8.6 \times 10^{-16}\ \Sigma(4.86\,
{\rm GHz}),\eqno(6)$$
where $\Phi({\rm H}\alpha )$ is the surface brightness of H$\alpha$
in erg~cm$^{-2}$~s$^{-1}$~arcsec$^{-2}$
and $\Sigma(4.86\,{\rm GHz})$ is the radio surface flux
density in $\mu$Jy~arcsec$^{-2}$.  The peak observed
$\Sigma(4.86\,{\rm GHz})$ is $\approx 11\,\mu$Jy
arcsec$^{-2}$ in the northwest sector of the nebula, which predicts
$\Phi({\rm H}\alpha )\ =\ 9.6 \times 10^{-15}$ erg~cm$^{-2}$~s$^{-1}$~arcsec$^{-2}$.
Even if we allow for the maximum extinction along the line of sight
(4.6 mag at H$\alpha$), we would still expect to see a surface flux of
$1.4 \times 10^{-16}$ erg~cm$^{-2}$~s$^{-1}$~arcsec$^{-2}$ at this location,
which should be a prominent feature in Figure~5.  Instead, there is no
sign of such emission to a limit at least one order of magnitude
smaller.  The absence of H$\alpha$ emission at this level is difficult
to understand if the radio emission is truly thermal bremsstrahlung
at $T \approx 8,000$~K.

The second failing of the H~II region hypothesis is the absence of
an exciting star within the radio shell.  The radio flux requires 
a B0 star of $10.3 \leq m_V \leq 15.8$.
We have ruled out spectroscopically the bright star of $R \simeq 13$ at
the bottom of Figure~3; it is an early K star.  The A star 
of $R = 16.8$ closest to the X-ray source is also inadequate, as are
half a dozen red stars of $R \approx 16-17$ which lie within the area
of the radio shell.  Thus, even after taking into account the
maximum extinction along this line of sight, it is not likely that
we have missed a star which could excite such an H~II region at any
distance less than 10~kpc.  And at a larger distance than this, one would
be suspicious of the presence of an H~II region more than 500~pc from
the Galactic plane.  We consider that an H~II region classification
of \vla\ is therefore ruled out by its high polarization and lack of 
corresponding optical evidence.  Since we also lack an optical counterpart
of \xray, we proceed to consider scenarios in which these sources
could be Galactic objects associated with each other
and powered by a neutron star.

\subsection {A Young Supernova Remnant?}

If \vla\ and \xray\ are associated, then it is highly likely
that they reside in the Galaxy.  We consider here whether
the radio shell can be a small-diameter supernova remnant (SNR) or
a pulsar wind nebula.  The SNR hypothesis encounters an
immediate difficulty, since shell components generally have steeper
radio spectra characterized by $\alpha = 0.4-0.7$, where
$S_{\nu} \propto \nu^{-\alpha}$.  The compact cores of plerionic
(Crab-like) remnants have flatter radio spectra ($\alpha = 0.0-0.3$),
but they have center-filled morphologies.  

The observed $\simeq 25\%$ linear polarization is high, but not unprecedented
in Galactic SNRs. The integrated polarization of the Crab Nebula,
for example, is $\sim 10\%$, but some features in the nebula show up
to 50\% linearly polarized flux (Wilson, Samarasinha, and Hogg 1985). The
Crab-like remnant G21.5--0.9 exhibits 20--30\% polarization in a circumferential
ring, although the integrated polarized fraction for the whole remnant is
somewhat lower (Becker \& Szymkowiak 1981). The typical polarized fraction for
shell-like remnants is 5--10\%. Extragalactic radio sources also exhibit
polarization, but typical values are a few percent, and no extended source with
a polarized fraction $>20$\% at 20 cm has ever been detected. Irrespective of
these comparisons, the high degree of linear polarization observed leads
to the inescapable conclusion that the radio source associated with
\mystery\ is nonthermal.

There is no clear manifestation of the shell at any energy above the
radio.  The X-rays could be dominated by a point source, and our search
for optical emission-line filaments, which was sensitive in the
velocity range $\pm 1,000$ km~s$^{-1}$, yielded null results as described above.
Nevertheless, it is clear that both the radio and
the X-ray sources are nonthermal.  The X-rays could be in part magnetospheric
synchrotron emission from a young neutron star, and in part a compact
plerion that is unresolved by the \asca\ SIS.  One way that a
flat radio spectrum
can be accommodated is if the characteristic synchrotron frequency of
the lowest energy electrons is comparable to or greater than the observing
frequency, with corresponding constraints on the magnetic field strength
and electron energies as described in \S 3.4 below.
$S_{\nu} \propto \nu^{+1/3}$ obtains in the low-frequency limit.
Thus, there is enough evidence to warrant a SNR interpretation for \vla,
albeit one with extreme properties.

\subsection {A Pulsar Bow-Shock Nebula?}

An alternative interpretation of the radio shell is nonthermal emission
from a shock between a relativistic pulsar wind and the surrounding 
interstellar medium (ISM).  Two scenarios can be considered for
the required power in the pulsar wind.  The first assumes that the
shock is expanding with velocity $v_s$, and is driven by the
difference in the interior and exterior pressure.
The second method assumes that
the radio shell is a bow shock which travels at the velocity of the
pulsar $v_p$, and whose apex is located where the pulsar wind pressure
equals $\rho_0\,v_p^2$.  Since the incomplete radio shell in fact resembles
a bow shock, we chose the latter scenario for consideration.
Thus, the spin-down power
$\dot E$ is related to the ambient ISM density
$\rho_0 (\approx 1.4\, n_{\rm H}\,m_p)$  and the apex distance $r_0$ via
$$\dot E\ =\ 4\pi\,r_0^2\,c\,\rho_0\,v_p^2\ =\ 
1.9\times 10^{38}\ \left ({n_{\rm H} \over 0.1}\right )
\left ({d \over 3\,{\rm kpc}}\right )^2 
\left ({v_p \over 100\,{\rm km\,s^{-1}}}\right )^2\ \ {\rm erg\,s^{-1}}.\eqno(7)$$
Here we have measured $1.\!^{\prime}7$ for the apex distance, and we assume
a relatively low-density ISM.  The rather large size of the radio shell
means that the required $\dot E$ is of order $10^{38}$ erg~s$^{-1}$
for any reasonable distance, pulsar velocity, or ISM density.

It is worth noting that the direction of the pulsar's velocity vector
under this hypothesis is nearly perpendicular to and away from
the Galactic plane,
consistent with its birth as a young star in the disk.  While not in itself
strong evidence that this is a young pulsar, a velocity in the
opposite direction would have been difficult to accommodate, as this
source is already at just about the maximum $z$-height
expected for a Population~I object ($z = 160$ pc for $d = 3$ kpc).

Under the bow shock hypothesis we assume that the radio flux is
nonthermal emission from the shocked pulsar wind itself -- i.e., the 
reverse shock rather than the shocked ISM -- as theorized by Hester \&
Kulkarni (1988) in their model of the core of the supernova
remnant CTB~80.   A Crab-like pulsar wind is thought to contain
relativistic electrons with $\gamma \approx 1 \times 10^6$
(Kennel \& Coroniti 1984a,b).  We hypothesize an
electron-positron wind, and we suppose that the shock produces a
relativistic Maxwellian distribution of particle energies
rather than a power law,
and an equipartition magnetic field of $B \approx 4 \times 10^{-5}$.
Then the synchrotron power (in $\nu F_{\nu}$) will be emitted mostly near
$\nu \approx 25\,\nu_c\, \approx 4 \times 10^{15}$~Hz (Tavani 1996)
where $\nu_c$ is the characteristic
frequency $\nu_c = (3 \gamma^2 e B)/(4\pi m_e c)$.
If a suprathermal power-law tail develops, then
the frequency of the synchrotron peak can move down a factor of 10
to $\nu \approx 2.5\,\nu_c\, \approx 4 \times 10^{14}$~Hz.
Thus, we may expect little X-ray emission coincident with the radio shell.  
The low-energy limit of the synchrotron emission from such a thermal
distribution has an asymptotic
spectrum of the form $F_{\nu} \propto \nu^{+1/3}$, just as in the
non-thermal case.

Assuming that the observed radio flux extrapolates
with $\alpha \approx 0.0$ up to $10^{16}$~Hz, 
the total synchrotron power from the bow shock is 
$\approx 8.6 \times 10^{36}$ erg~s$^{-1}$,
which is a significant fraction of the $\dot E$
inferred from the radius of the shell.   We would
not easily detect the optical synchrotron emission from the reverse
shock because its surface brightness would be at most
$\approx 11\,\mu$Jy arcsec$^{-2}$ as it is in the radio, which
corresponds to 21.1 mag arcsec$^{-2}$ in the $R$ band, or
25.7 mag arcsec$^{-2}$ after correcting for the maximum extinction.
Any steepening of the spectral index above the radio band would
further decrease this optical estimate.

The majority of $\dot E$ may go into accelerating and heating the shocked ISM.
Although the forward shock from such pulsar winds is sometimes
detected in H$\alpha$, the low density of the ISM in this case,
and the relative inefficiency of the non-radiative shock which is usually 
inferred, would predict much less H$\alpha$ emission than was
required in the H~II region scenario for \vla.  A simple
scaling relation can be derived to approximate the average
surface brightness in H$\alpha$ of a nonradiative shock.
According to Raymond (1991), it is the neutral fraction
of hydrogen passing freely through the shock that emits on average
0.2 H$\alpha$ photons per atom via collisional excitation before
it is ionized by the hot, shocked ISM.  This number is relatively
independent of shock velocity because the collisional excitation
and collisional ionization cross sections scale similarly with
temperature.  If the bulk of the H$\alpha$ is emitted in a
hemispherical region of radius $r_0$, then the average H$\alpha$
surface brightness is
$$\Phi({\rm H}\alpha)\ =\ 1.1 \times 10^{-18}\ 
\left ({n_{\rm H} \over 0.1}\right )\ 
\left ({v_p \over 100\,{\rm km\,s^{-1}}}\right )\ \
{\rm erg\ cm^{-2}\ s^{-1}\ arcsec^{-2}}. \eqno(8)$$
After including the effects of extinction, this value is
2--3 orders of magnitude smaller than the limiting
sensitivity of our H$\alpha$ image.

In this scenario, neither the reverse shock nor the forward shock are
energetic enough to emit the hard X-rays which are seen from \xray.
Since the X-rays are apparently more compact in size than the radio
shell, we would assume that they are entirely magnetospheric
synchrotron emission from a young neutron star and/or a compact
plerion that is unresolved by the \asca\ SIS.  Its power-law spectral
index $\Gamma = 1.5$ is compatible with this interpretation, as is its
lack of variability.  The 2--10 keV luminosity
of this source is $1.7 \times 10^{33}\ (d/3\,{\rm kpc})^2$ erg~s$^{-1}$,
which is $\approx 10^{-5}$ of the pulsar spin-down power as inferred from
the bow-shock interpretation of the radio shell.  This is perhaps one
of the weaknesses of this model, since the 
ratio $L_x/\dot E$ for rotation-powered pulsars is
typically observed to be $10^{-4}-10^{-3}$.
While $\dot E$ can perhaps be reduced by lowering
the ambient density $n_{\rm H}$ to 0.01, the unusual flat radio spectrum
of \vla\ remains a significant mystery.  Thus, none of the scenarios
that we have considered is without difficulty.

\subsection {A Pulsar Wind Bubble?}

Some of the problems associated with the large spin-down power
inferred from the bow-shock interpretation of the
radio source \vla\ could be ameliorated if it is a bubble confined by
the static pressure $nkT$ of the ISM instead of the dynamic pressure
$\rho_0\,v_p^2$.  This scenario would require the pulsar to have fortuitously
a smaller velocity than the thermal speed of the ISM, but if it did, the
inferred spin-down power could be much smaller, 
$$\dot E\ =\ 4\pi\,r_0^2\,c\,(n_e+n_i)\,k\,T\ =\ 
2.4\times 10^{36}\ \left ({n_{\rm H} \over 0.1}\right )
\left ({d \over 3\,{\rm kpc}}\right )^2 
\left ({T \over 10^4\,{\rm K}}\right )\ \ {\rm erg\,s^{-1}}.\eqno(9)$$
We assume an ambient temperature of $\sim 10^4$~K as appropriate if the
ISM were pre-ionized by the recent UV flash from the supernova which produced
the neutron star, or if an older pulsar is now surrounded by the cooling
remnant of the past supernova shock.
In this case, the X-ray luminosity could indeed be $10^{-4}-10^{-3}$
of $\dot E$ as is observed from most other pulsars.
We would still not expect to detect diffuse optical
continuum or H$\alpha$ emission associated with the radio shell,
although the extrapolated radio spectrum of the shell would have
to turn down below $\sim 10^{15}$~Hz in order not to exceed $\dot E$.
The pulsar would then be required to have a very high efficiency
of conversion of $\dot E$ into $\gamma$-rays if it were to be the counterpart
of the EGRET source \source.  The observed photon flux of \source\ above
100~MeV, $4.1 \times 10^{-7}$ photon~cm$^{-2}$~s$^{-1}$ translates, for
the photon spectral index of 2.14, into a luminosity of
$3.7 \times 10^{35}\,(d/3\,{\rm kpc})^2$ erg~s$^{-1}$,
approximately 15\% of $\dot E$ from Equation (9).  Such high $\gamma$-ray
efficiencies are also inferred for the adolescent
pulsars Geminga and PSR~B1055--52; they could be reduced
by a modest amount of anisotropy in the $\gamma$-ray beam pattern.
It could be argued that the shape of the radio source \vla\
resembles an incomplete spherical bubble as much as it
does a cometary bow-shock surface.

\section {Implications of the Identification (or Not) of \source\ }

Despite the lack of a completely satisfactory theory linking
the X-ray source \xray\ and the radio shell \vla, we are
encouraged that their properties are not inconsistent with
what we would expect for a pulsar counterpart of \source.
Of all the pulsars detected by EGRET, the Crab is the only one
whose $\gamma$-ray spectral index ($\Gamma = 2.19 \pm 0.02$)
is identical to that of \source\ ($\Gamma = 2.24 \pm 0.14$).
All of the less energetic pulsars have flatter spectra,
and there is a clear trend of spectral flattening 
(and increasing $\gamma$-ray efficiency) with increasing
age or decreasing spin-down power.
Furthermore, since the observed $\gamma$-ray flux of the Crab is
approximately 5 times that of \source, the estimated
distance to \xray\ of 3~kpc, 1.5 times that of the Crab,
is basically what one would expect for a pulsar of half the
spin-down power and similar $\gamma$-ray efficiency.
At this intermediate Galactic latitude
of $3^{\circ}$, a distance of 3~kpc is just about the maximum
that one would expect for a young Population I object which should
be found within the ISM layer of the disk, and it is just about right
to accommodate the observed X-ray absorption column of
$6.3 \times 10^{21}$~cm$^{-2}$.  Essentially, by selecting an EGRET
source at this Galactic latitude, we are predisposed to finding
a pulsar that is similar to the Crab but more distant.
Accordingly, it is perhaps not surprising that the bow-shock
interpretation of the radio shell \vla\ requires 
$\dot E \sim 2 \times 10^{38}$ erg~s$^{-1}$, about half that
of the Crab.  If this identification for \source\ is correct,
the main difference between it and the Crab would be the
presence of a prominent bow shock, and the relative
weakness of X-ray emission from the pulsar magnetosphere
and/or plerion.  The failure to detect X-ray pulsations 
from \xray\ is mysterious, however, 
unless a compact plerion dominates over the pulsar emission.

We presented an alternative scenario in which the spin-down power
of the putative pulsar could be much lower, $\approx 2.4 \times 10^{36}$
erg~s$^{-1}$.  Although this would allow an older and therefore
more ``common'' class of pulsar, the 3~kpc distance
would strain its $\gamma$-ray 
production efficiency if it behaves similarly to $\gamma$-ray
pulsars in this regime of $\dot E$,
such as Vela, PSR~B1951+32, and PSR~B1706--44.
Those three pulsars have (isotropic) $\gamma$-ray efficiencies between
0.4\% and 7\%, while a $\gamma$-ray luminosity of
$3.7 \times 10^{35}\,(d/3\,{\rm kpc})^2$ erg~s$^{-1}$
for \source\ would require an efficiency of 15\%.  Perhaps this is
not implausible, however, because the older pulsars
Geminga and PSR~B1055--52 have even higher apparent efficiency.
Thus, it is possible that we are dealing with a highly efficient example
of an intermediate age $\gamma$-ray pulsar.

The observed X-ray spectral index of \xray, $\Gamma = 1.5$,
was in fact predicted
for strong $\gamma$-ray pulsars by Wang et al. (1998)
in the context of the outer-gap model.  In this model,
an outer-gap accelerator sends $e^{\pm}$ pairs flowing inward and outward
along open magnetic field lines.  These particles continuously
radiate $\gamma$-rays by the curvature mechanism.  When the
inward flowing particles approach the surface of the star,
the $>100$~MeV $\gamma$-rays that they emit convert
into secondary $e^{\pm}$ pairs in the inner magnetosphere
wherever $B\,{\rm sin}\phi > 2 \times 10^{10}$~G, where $\phi$ is the
angle between the photon and the {\bf B} field.  Those
secondary pairs must radiate away their energy instantaneously in the
strong local {\bf B} field.  Such a
synchrotron decay spectrum has $\Gamma = 1.5$ between $E_{\rm min} = 0.1$~keV
and $E_{\rm max} = 5$~MeV.
We emphasize that in this theory,
the X-ray power-law is {\it not} supposed to be a simple
continuation of the EGRET spectrum.  Rather, it is a separate component
radiated by the secondary $e^{\pm}$ pairs that are created when
some of the primary $\gamma$-rays convert in the strong {\bf B} field.
The X-ray luminosity generated by this mechanism is
$$L_x(2-10\ {\rm keV}) \approx 1.5 \times 10^{31}\ f\
\left ({0.1 \over P}\right )^2\,\left ({B_{\rm p} \over 10^{12}}\right )^{1/2}
\ \ {\rm erg\ s^{-1}}, \eqno(10)$$
where $P$ is the rotation period, $B_{\rm p}$ is the
surface polar magnetic field, and $f$ is the fraction of the
Goldreich-Julian current that flows through the outer-gap
accelerator back to the polar cap.  This mechanism could make a
significant contribution to the observed X-ray luminosity of \xray\ only 
if $P$ is small, $\sim 0.01$~s, {\it and\/} if $f$ is of order unity.
Otherwise, most of the X-rays must be generated by another mechanism,
including perhaps an extended synchrotron nebula.

An alternative hypothesis,
that \xray\ is a twin of the intermediate-age pulsar Geminga,
would
be difficult to accommodate since Geminga's $\dot E = 3.3 \times 10^{34}$
erg~s$^{-1}$ is inadequate to account for the $\gamma$-ray luminosity
of \source\ at a distance of 3~kpc unless a beaming factor $> 10$
is supposed.  Also, the nonthermal X-ray luminosity of \xray,
$1.7 \times 10^{33}\ (d/3\,{\rm kpc})^2$ erg~s$^{-1}$,
is 3 orders of magnitude more than that of Geminga. 

If \xray\ proves {\it not} to be the identification of \source,
then the absence of any other X-ray candidate at the level of
$\sim 4 \times 10^{-14}$ erg~s$^{-1}$ is difficult to reconcile
with any of the established classes of $\gamma$-ray emitters.
If \source\ is a pulsar but \xray\ is {\it not}
its counterpart, it implies that highly efficient
(or highly beamed) $\gamma$-ray pulsars can avoid producing 
soft or hard X-rays at a level below $10^{-4}$ of their apparent
$\gamma$-ray luminosity.  At least two mechanisms of X-ray
emission have been observed to accompany all $\gamma$-ray
pulsars at such levels or higher, and they
were explained in the context of the outer-gap model
by Wang et al. (1998).  One is the 
synchrotron emission from secondary pairs as described above.
The second is thermal emission
from the heated polar caps that
are impacted by the inward-going accelerated particles
from the outer-gap accelerator, as well as from
the original heat of formation emitted over the whole surface. 
If \source\ were a low-luminosity pulsar like Geminga,
then it must be less than 1~kpc distant in order to be detected
by EGRET, yet the intervening column density must be greater than
$10^{21}$~cm$^{-2}$ in order for its thermal soft X-rays not to
reach us.  Many of the unidentified EGRET sources could be similarly
situated radio-quiet pulsars, made exceedingly difficult to identify
in X-rays because of interstellar absorption.

One might still hypothesize that \source\ is an extragalactic
object.  The one tempting piece of evidence along those lines is
the steep $\gamma$-ray spectral index, which if it is not due to
a Crab-like pulsar, is compatible with many of the known EGRET blazars.
But as discussed above, there is no compact, flat-spectrum radio source
in this field to an upper limit $\approx 20$~mJy, or 50--100 times fainter
than the typical identified EGRET blazars.  Furthermore, there is no
extragalactic X-ray source in this field other than possibly
\xray\ itself, and it is radio quiet (as a compact source) at the
1~mJy level.  Thus, there is no escaping the fact that \source\
would be a member of a new class of $\gamma$-ray source if extragalactic.
However, it would not necessarily be unique in this regard.
Similar considerations concerning the high-latitude source
3EG~J1835+5918 were recently presented by Mirabal et al. (2000),
where even more stringent upper limits to its potential pulsar or
blazar counterparts were obtained.  Since many EGRET sources remain
unidentified, more of them may prove, upon detailed study and especially
after more precise localizations are obtained by {\it GLAST},
to be of a previously unrecognized type, for example, the 
frequently imagined ``radio-quiet blazar''.
It is of interest that several $\gamma$-ray loud quasars are seen
to have flat X-ray spectra in the 2--10 keV band, with $\Gamma = 1.3-1.5$
(Tavecchio et al. 2000).  Their broad-band spectra can be fitted with
Inverse Compton jet models only if a significant proton component or
Poynting flux outside the emission region is the main carrier of power.
If \xray\ were a radio-quiet blazar of this type, then its flat X-ray
spectrum and lack of an optical counterpart might be just what is
needed for the prototype of a new or extreme class of $\gamma$-ray
loud AGN.

\section {Conclusions}

\ro\ and \asca\ observations of the error circle of \source\
reveal only one candidate X-ray counterpart with an intrinsic
2--10~keV flux of $1.56 \times 10^{-12}$ erg~cm$^{-2}$~s$^{-1}$,
fitted by a power-law spectrum of photon index
$\Gamma = 1.51 \pm 0.14$ and
$N_{\rm H} = (6.3 \pm 1.3) \times 10^{21}$~cm$^{-2}$.
There is no other candidate to a flux limit
of $\sim 6 \times 10^{-14}$ erg~cm$^{-2}$~s$^{-1}$.  This
X-ray source happens to coincide with a highly polarized, incomplete
radio shell which resembles a bow-shock nebula or a wind-blown
bubble.  The X-ray measured column density implies a distance
of $\approx 3$~kpc, although an extragalactic source has not
been ruled out definitively.  The flat radio spectrum is
different from that of all other shell SNRs, and
our optical searches have turned up no evidence for associated
emission-line gas.  On balance, we favor the
hypothesis that \xray\ is indeed an energetic
pulsar counterpart of \source,
and that \vla\ is an associated nebula
powered by a large fraction of the pulsar spin-down power.
The best method for proving this and quantifying the age 
and energetics of this unique source would be
to obtain more sensitive X-ray imaging and timing observations.
We have planned both {\it Chandra} and {\it XMM} observations.
{\it Chandra} has the ability to resolve a point source from
the hypothesized compact synchrotron nebula, and to obtain a precise
position for deeper optical follow-up.  {\it XMM} with its high
throughput may permit a pulsar discovery
and a precise ephemeris to be developed, which could
then lead to detection of pulses in the archival \asca\ and
EGRET data, and in future observations by  {\it GLAST}.  
Deeper optical spectroscopy, and possibly infrared imaging,
would constitute a more definitive test for an extragalactic counterpart
of \xray\ such as a Seyfert galaxy
or a quasar.  The absence of such a counterpart would itself be strong
evidence that \xray\ is a neutron star.

\acknowledgments{
This work was supported by NASA grants NAG 5-3229 and NAG 5-7814.
We thank R. H. Becker for obtaining an optical spectrum with the
Keck telescope, and M. Eracleous for assistance with other spectroscopic
observations at Kitt Peak.  We also thank R. H. Becker for calling
our attention to the important polarization information contained
in the NVSS images.}

\clearpage

\begin{deluxetable}{rlcrrlrrl}
\tablenum{1}
\tablecolumns{9}
\tablewidth{0pc}
\tablecaption{\ro\ HRI X-ray Sources in the Field of \source\ }
\tablehead
{
X-ray & Position & & & Optical & Position\\
R.A. & Decl. & Unc.\tablenotemark{a} & Counts & R.A. & Decl. & \omit\hfil $R$ \hfil & \omit\hfil $B$ \hfil & ID \\
(h \hskip 0.7em m \hskip 0.7em s) & ($\circ\ \ \ \prime\ \ \ \prime\prime$) & ($\prime\prime$) & (ksec$^{-1}$) &
(h \hskip 0.7em m \hskip 0.7em s) & ($\circ\ \ \ \prime\ \ \ \prime\prime$) & \omit\hfil (mag) \hfil & \omit \hfil (mag) \hfil \\
}
\startdata
22 24 30.25 & +61 28 19.0 & $\pm 2.0$ & $ 2.6 \pm 0.9$ & 22 24 30.39 & +61 28 19.3 & 14.9 & 17.0 & dMe star \\
22 26 28.26 & +61 45 57.8 & $\pm 2.4$ & $ 1.6 \pm 0.6$ & 22 26 28.30 & +61 45 55.5 & 14.3 & 16.0 & M star \\
22 26 38.93 & +61 13 30.2 & $\pm 3.8$ & $ 1.6 \pm 0.6$ & 22 26 38.79 & +61 13 31.4 & 13.2 & 14.8 & Herbig Ae/Be \\
22 29 04.97 & +61 14 12.9 & $\pm 2.7$ & $ 2.5 \pm 0.6$ & --------------- & --------------- & $\geq 21.3$ & $>24.0$ & ------ \\
22 29 29.09 & +61 04 53.6 & $\pm 1.4$ & $ 2.3 \pm 0.5$ & 22 29 29.23 & +61 04 52.2 & 12.9 & 13.9 & K star \\
22 30 49.68 & +61 30 52.5 & $\pm 1.1$ & $13.1 \pm 1.2$ & 22 30 49.32 & +61 30 51.9 & 12.8 & 13.9 & K star \\
\enddata
\tablenotetext{a}{ 90\% confidence uncertainty in each
coordinate.}
\end{deluxetable}
\clearpage

\plotone{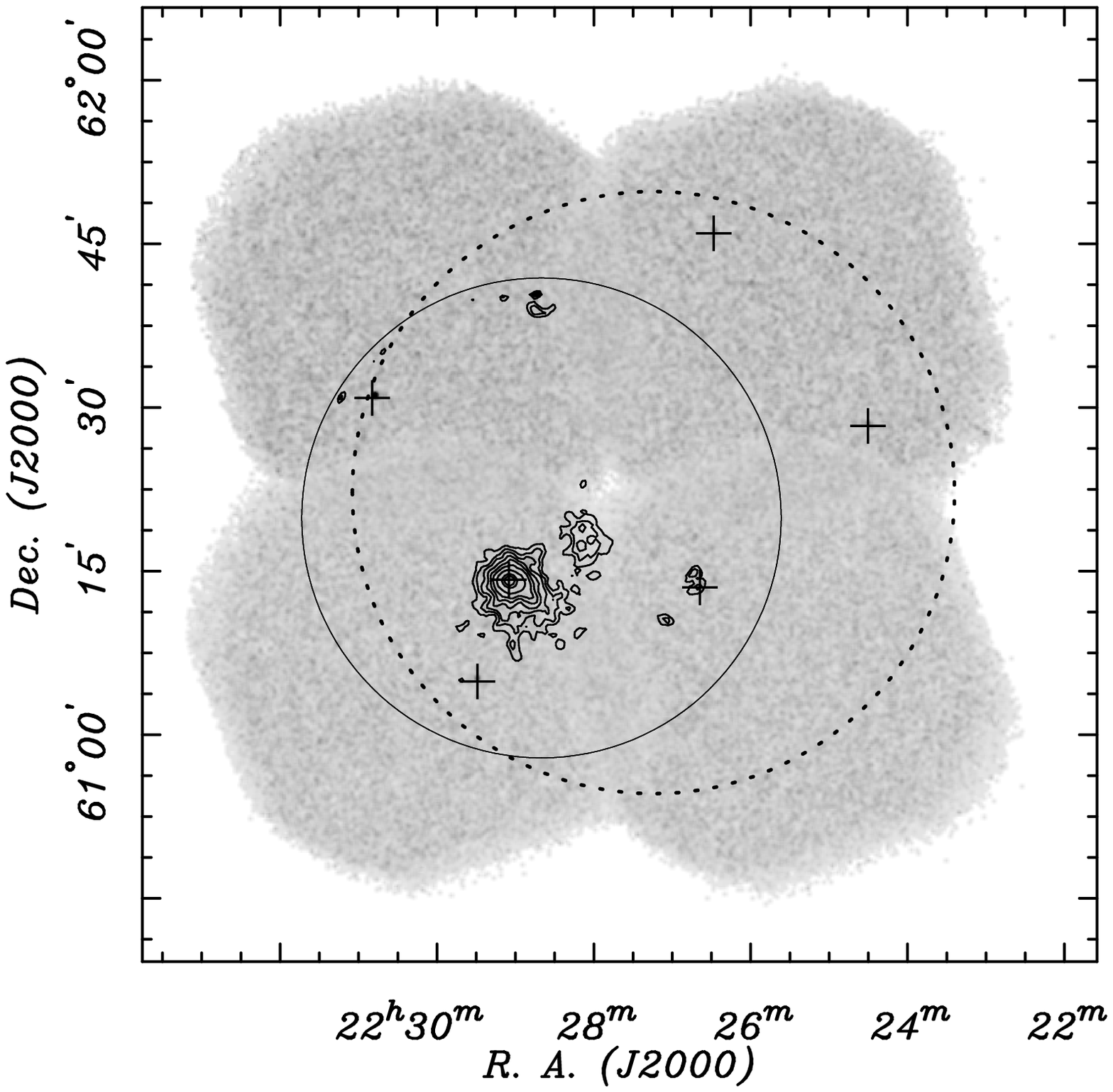}
\vskip 0.5 truein
\figcaption[fig_1.eps]{A montage of four {\it ROSAT\/} HRI images covering
the 95\% error circle of 3EG~J2227+6122 ({\it dashed circle}).
There are six {\it plus signs} at the locations of HRI sources
(also listed in Table~1);
all but one of these are bright stars.
The {\it solid circle} and the contours within it are the {\it ASCA} GIS
images, showing a bright, hard source coincident with the
only unidentified HRI source.\hfill}

\plotone{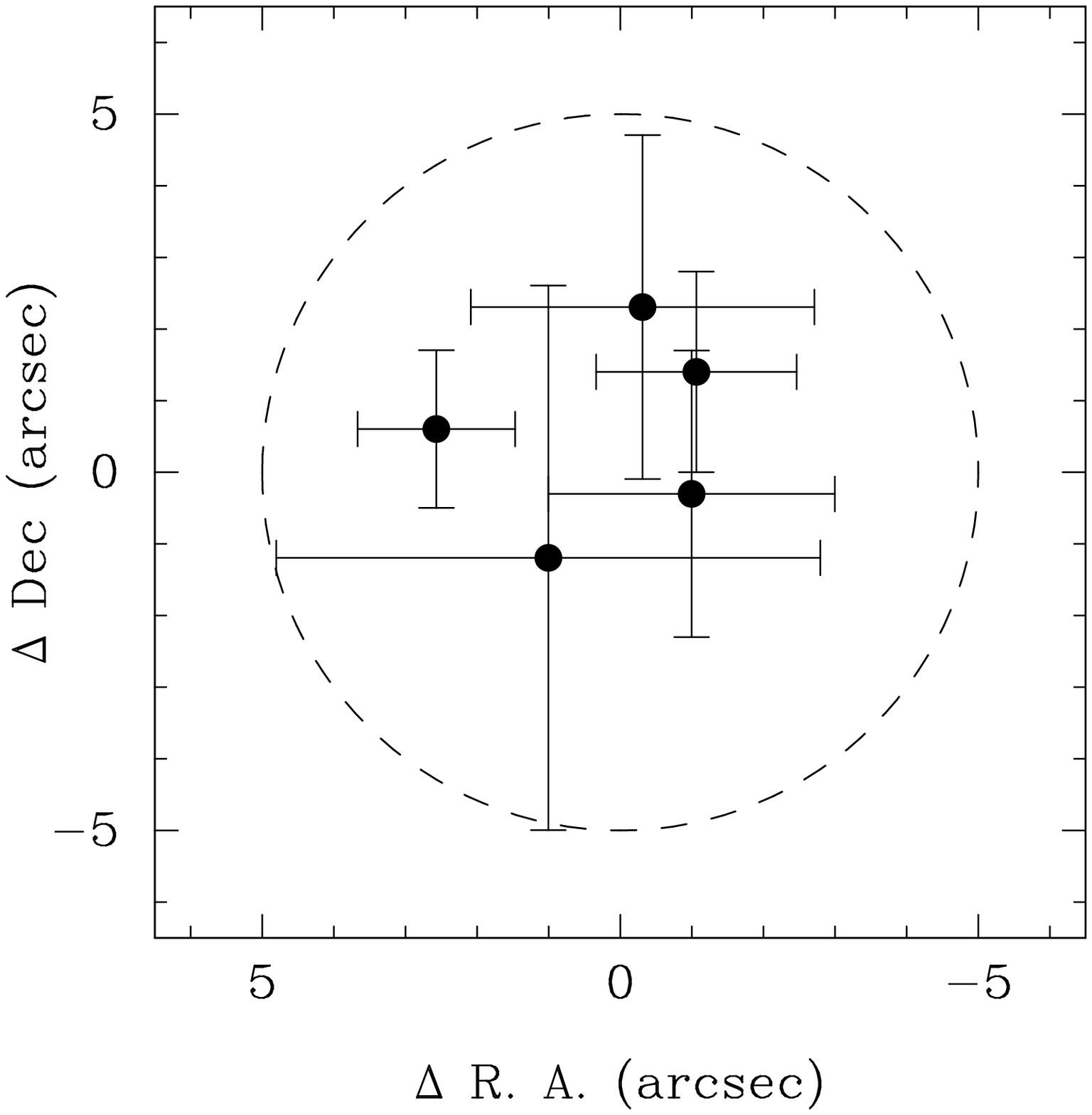}
\figcaption[fig_2.eps]{
\ro\ HRI X-ray source positions and their 90\% confidence 
uncertainties compared with the optical positions of the five bright stars
with which they are identified.  The offset between the X-ray and
optical position is plotted.  The illustrated
$5^{\prime\prime}$ radius circle
is, therefore, a conservative error estimate on the position of the
unidentified source \xray. \hfill}

\plotone{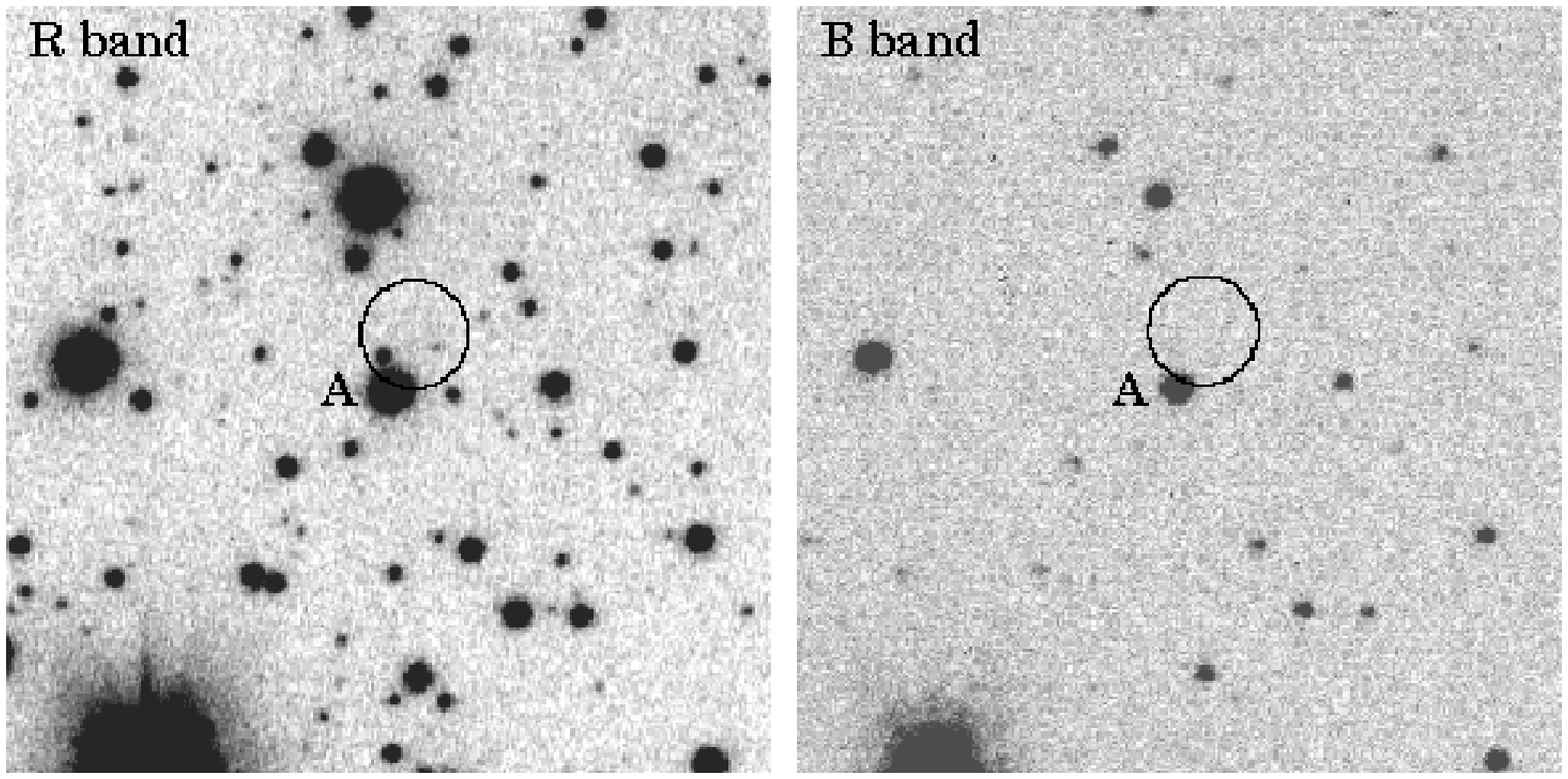}
\vskip 0.5 truein
\figcaption[fig_3.eps]{
Optical images of the location of the unidentified X-ray source
\mystery\ from the MDM 2.4m telescope.
The error circle is $5^{\prime\prime}$ in radius.
Limiting magnitudes are 24.5 in $R$
and 24.0 in $B$. Star A is a highly reddened A star, and the $R = 21.3$
object just north of it is either a star or a galaxy with no emission lines.
The faintest detected object within the error circle has $R = 23.00 \pm 0.10$.
\hfill}

\epsscale{1.3}
\vbox{
\vskip -1.0 truein
\plotone{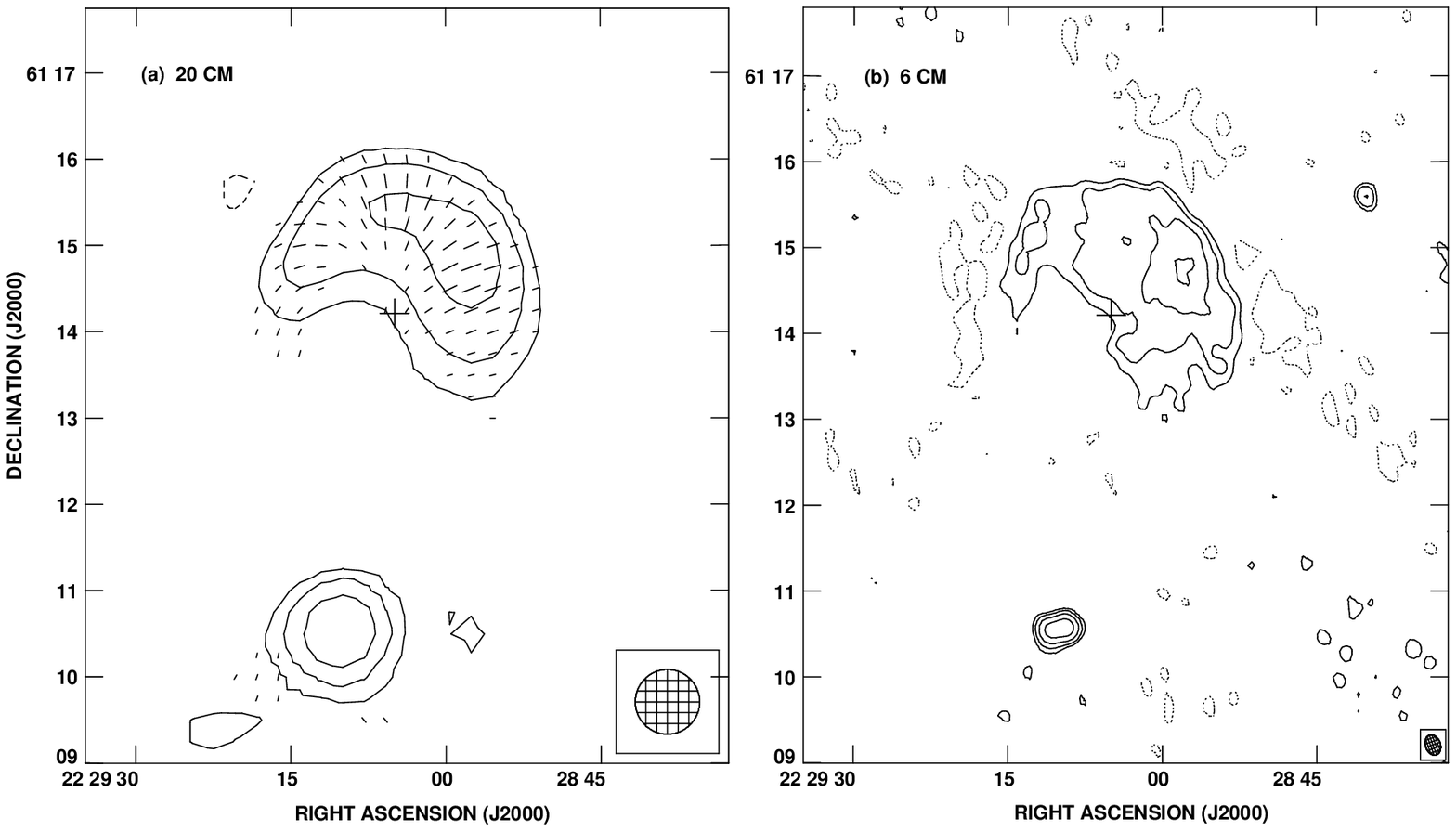}
\vskip -4.0 truein}
\epsscale{1.}
\figcaption[fig_4.eps]{({\it a}) The 20~cm NVSS map showing the shell-like radio source
\vla\ with polarization vectors superposed.  The {\it plus sign} is the
location of the X-ray source \mystery.  Contour levels are
2.2, 4.4, and 8.8 mJy beam$^{-1}$.
The length of the polarization vector corresponds to 0.30 mJy beam$^{-1}$
arcsec$^{-1}$. The hatched circle at the lower right represents the beam size.
({\it b}) A 6~cm VLA map obtained in the D configuration.  The beam is
a factor of 3 smaller than in {\it a}.
Contour levels are --0.15, 0.15, 0.3, 0.6, and 1.2 mJy beam$^{-1}$.
The resolved radio source below the
shell has a steep spectrum and is presumably a background
AGN.  This background radio source is not detected in X-rays.\hfill}

\plotone{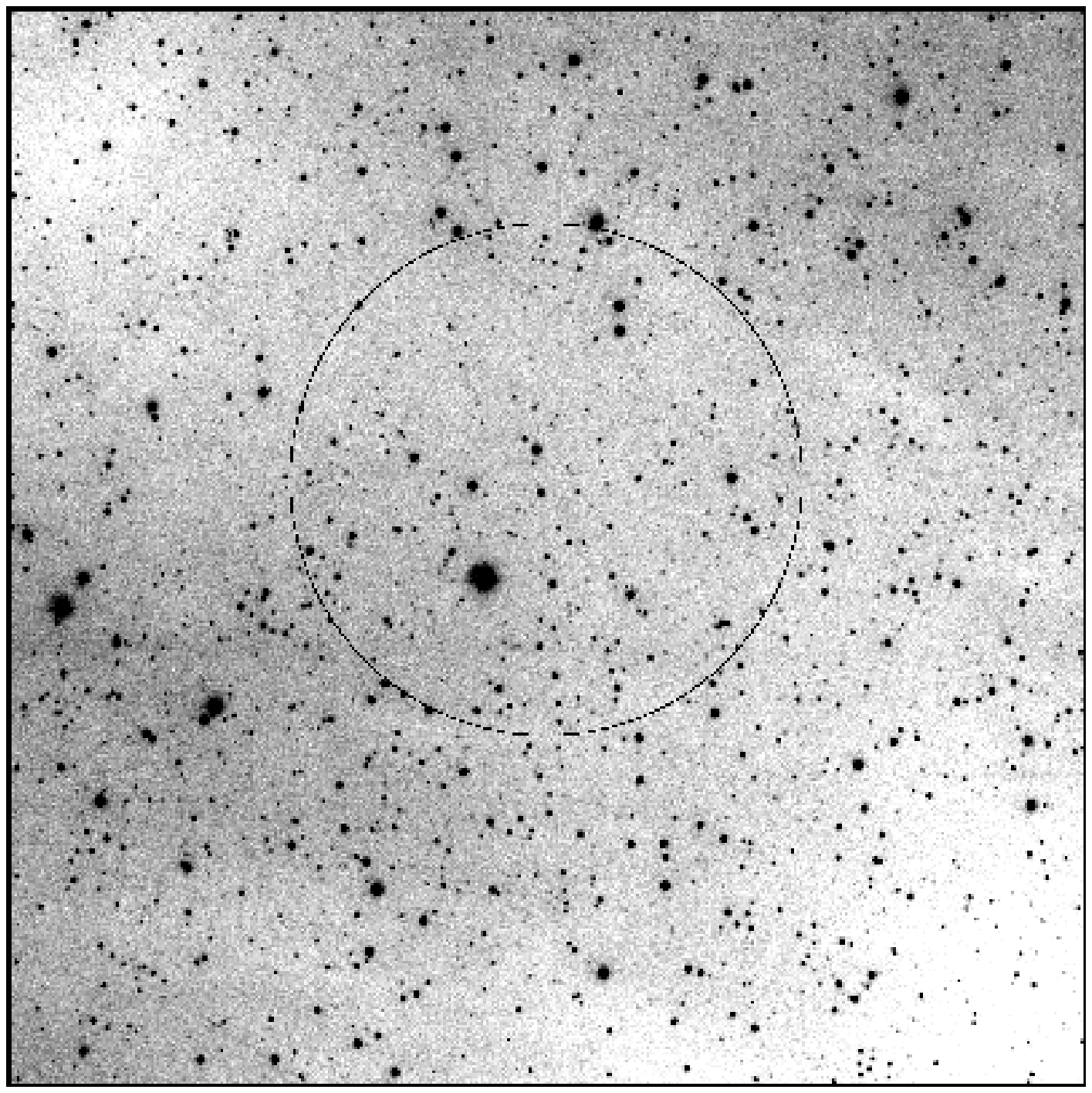}
\figcaption[fig_5.eps]{An H$\alpha$ image from the MDM 2.4m telescope
of a $7.\!^{\prime}3 \times 7.\!^{\prime}3$ region around 
the radio shell \vla.
A 39~\AA\ wide filter was used, and the brightest regions have
a surface brightness of $1.7 \times 10^{-16}$
erg~cm$^{-2}$~s$^{-1}~{\rm arcsec}^{-2}$.
The circle marks the approximate boundary of the radio shell as seen
in Figure 4.
There is apparently no H$\alpha$ emission associated with it.\hfill}

\plotone{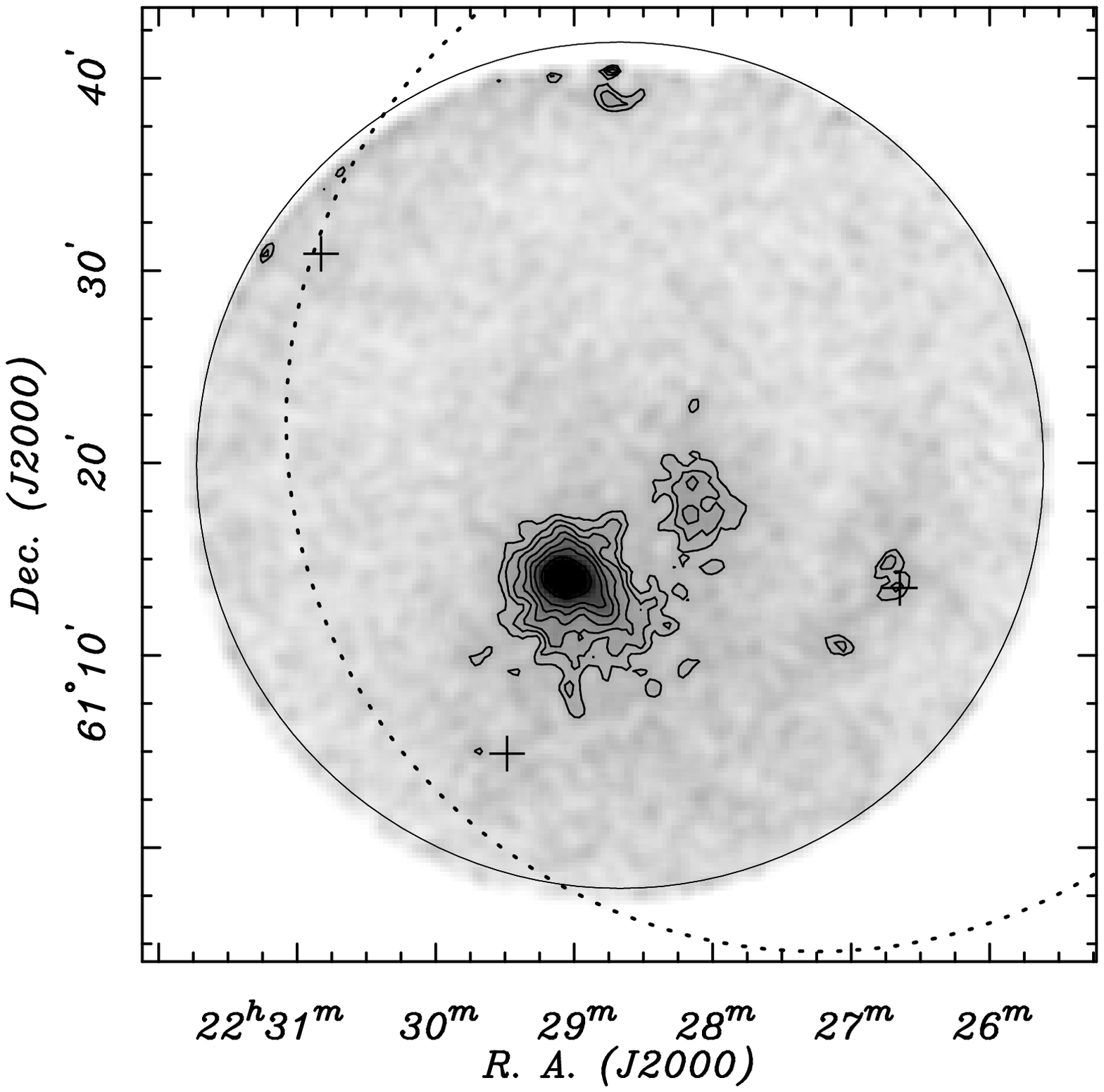}
\vskip 0.5 truein
\figcaption[fig_6.eps]{The \asca\ GIS images in the 0.8--8.0~keV band,
with the locations of \ro\ HRI sources
({\it plus signs}) superposed.  Contours are used to highlight
the grey scale.
Diffuse, softer X-ray emission is evident
surrounding the hard source \xray\ and to the northwest of
it.  The weak source far to the west is coincident with 
the Herbig Ae/Be star.
\hfill}

\plotone{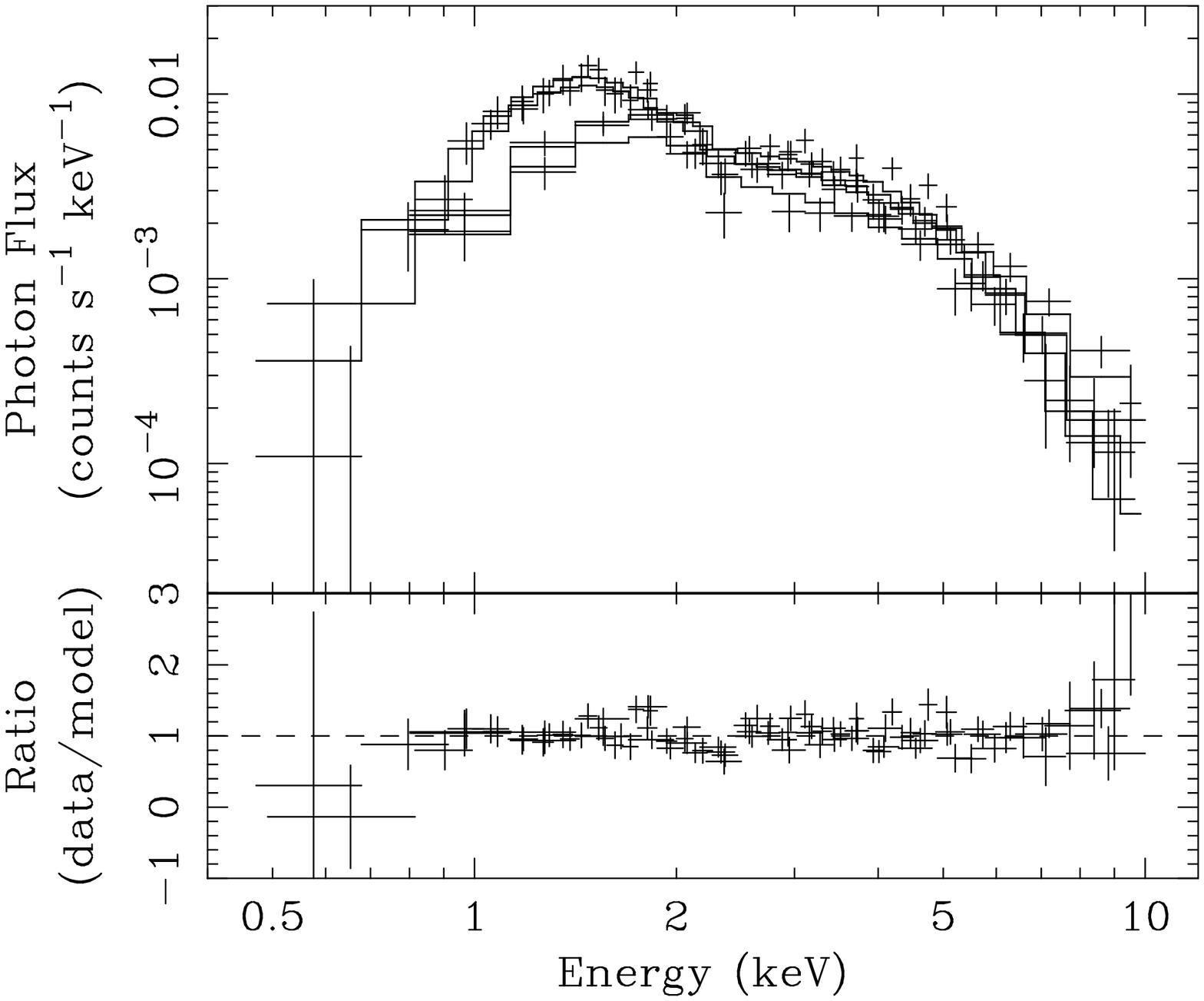}
\figcaption[fig_7.eps]{\asca\ SIS and GIS spectra of \mystery, and residuals
from a power-law fit.\hfill}

\plotone{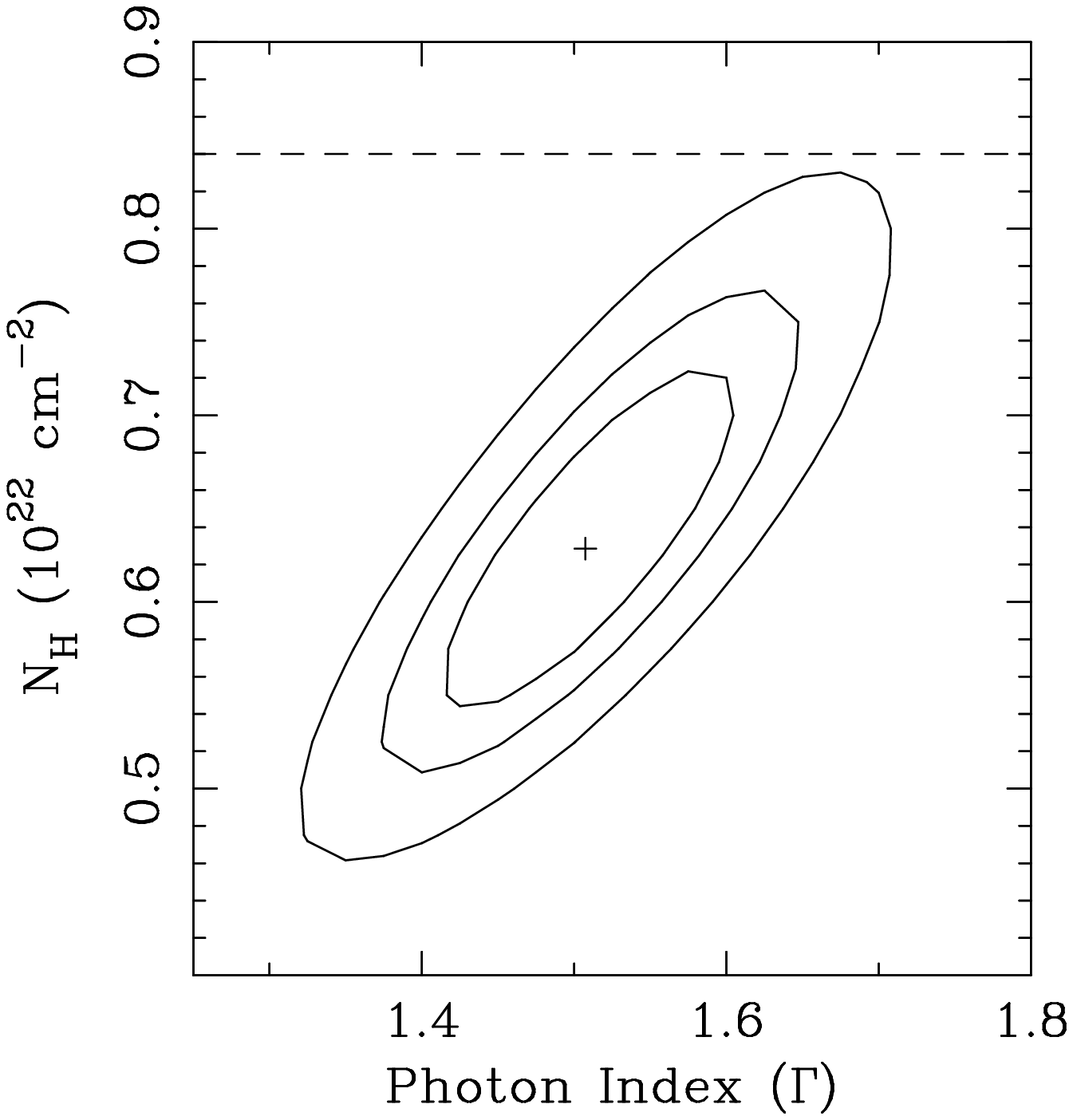}
\vskip 1.0 truein
\figcaption[fig_8.eps]{Confidence contours of parameters fitted to the
\asca\ spectrum of \xray.  Confidence limits are 68\%, 90\%,
and 99\% for two interesting parameters.  The total Galactic
21~cm column density in this direction is indicated by a dashed line.
\hfill}

\end{document}